\documentclass[runningheads, envcountsame]{llncs}
\usepackage[T1]{fontenc}
\usepackage{graphicx}

\usepackage{amssymb}
\usepackage{amsmath}
\usepackage{cleveref}
\usepackage{algorithmicx}
\usepackage[Algorithm,ruled]{algorithm}
\usepackage{algpseudocode}
\usepackage{tikz}
\usetikzlibrary{decorations.pathmorphing, decorations.pathreplacing, decorations.shapes}

\usepackage[textsize=footnotesize,color=green!40]{todonotes}

\begin{document}
\title{Graph Exploration with Edge Weight Estimates}

\author{Matthias Gehnen\inst{1} \orcidID{0000-0001-9595-2992},
 Ralf Klasing\inst{2} \\
 and \'{E}mile Naquin\inst{3} \orcidID{0009-0009-1496-3717}
}

\authorrunning{M.~Gehnen et al.}

\institute{RWTH Aachen, Aachen, Germany \\
	\email{gehnen@cs.rwth-aachen.de} \and
CNRS, LaBRI, Université de Bordeaux, Talence, France \\%
 \email{ralf.klasing@labri.fr} \and
LaBRI, Université de Bordeaux, Talence, France\\ and University of Augsburg, Augsburg, Germany\\ 
 \email{emile.naquin@labri.fr}}

\maketitle  
\vspace*{-5mm}
\begin{abstract}
In the \textsc{Travelling Salesman Problem}, every vertex of an edge-weighted 
graph has to be visited by an agent who traverses the edges of the graph.
In this problem, it is usually assumed that the costs of each edge are given in advance,
making it computationally hard but possible to calculate an optimal tour for the agent.

Also in the \textsc{Graph Exploration Problem}, every vertex of a given graph must be visited, but here the graph is not known in the beginning - at every point, an algorithm only knows about the already visited vertices and their neighbors.

Both however are not necessarily realistic settings: Usually the structure of the graph 
(for example underlying road network) is known in advance, but the details are not.
One usually has a prediction of how long it takes to traverse through a particular road,
but due to road conditions or imprecise maps the agent might realize that a road will take slightly longer
than expected when arriving on it.
To deal with those deviations, it is natural to assume that the agent is able to adapt to the situation:
When realizing that taking a particular road is more expensive than expected, recalculating the 
tour and taking another road instead is possible.

In a sense, this setting lies in between the offline travelling salesman problem, and the
online problem of graph exploration, as it allows some computation at the beginning, but
remains an online problem during traversal through the graph.

We analyze the competitive ratio of this problem based on the perturbation factor $\alpha$ of the 
edge weights. For general graphs we show that for realistic factors smaller than $2$ there is no strategy that achieves a competitive ratio better than $\alpha$, which can be matched by a simple algorithm.

In addition, we prove an algorithm which has a competitive ratio of $\frac{1+\alpha}{2}$ for restricted graph classes like complete graphs with uniform announced edge weights. Here, we present a matching lower bound as well, proving that the strategy for those graph classes is best possible. 
 We conclude with a remark about special graph classes like cycles.

\end{abstract}


\section{Introduction}
When planning a tour for a travelling salesman, usually it is assumed that the distances between each two points are known in advance. Therefore, computing an optimal route is possible, although the computation might be hard. However, this assumption might not be realistic in real-world applications. Take a simple example:

Imagine a tourist who wants to visit $n$ sights across a city. The tourist is aware of the street map, and therefore knows the structure of the city he wants to explore. However, the time that it will take to travel on each road is not exactly known in advance, since unexpected traffic, accidents, or road conditions are not depicted in the tourist's street map. As soon as the tourist arrives at some street, he will get aware of the actual conditions and therefore of the time traversing the road takes.
 
We model this by assuming that the agent who explores the graph knows the graph structure together with an interval for each edge weight in advance. The actual, precise weight of each edge gets revealed when the agent is on a vertex incident to the edge. As we focus on the impact of the missing information, we assume that the agent has unbounded computational power and space.

In this article, we start with a formal definition of the graph exploration problem with edge weight estimates. Afterwards, we will give a short overview about the travelling salesman and graph exploration problem and see how the topic of this work lies in between them.

In Section~\ref{chapter-general-case}, we will see that in general it is not possible to hope for algorithms that achieve a better competitive ratio than the ratio between the upper and lower bound of the edge weight announcement. This also shows that a simple algorithm, that just calculates the shortest tour based on the lower bounds announced is best possible in some cases.

This however is not true for all graph classes, as it is shown in Section~\ref{chapter-special-case}: We will see that graphs like complete graphs or complete balanced bipartite graphs allow a competitive ratio of $\frac{1+\alpha}{2}$, when each edge weight is announced uniformly in an interval of $[1, \alpha]$. This can be achieved by using an algorithm that recalculates the cheapest possible tour in each step when new edge weights are revealed. We further show that this competitive ratio is best possible for those graph classes, but also that this algorithm is not able to calculate a better competitive ratio than $\alpha$ for other graph classes, even if they have uniform announced edge weights.

We finally conclude with an outlook to other special graph classes like cycles, and close with some questions that imposed while developing this framework.
\newpage
\subsection{Formal Definition of the Estimates Model}

In this article, we consider the travelling salesman problem as the underlying offline problem of our semi-online setting:
\begin{definition}[Travelling Salesman Problem (TSP)]
	Given a graph $G=(V,E)$ with weights $w(e)$ for each edge $e\in E$, a vertex $s \in V$ as a start-vertex and a vertex $t \in V$ as end-vertex, the \emph{Travelling Salesman (Path) Problem} is the problem of finding the shortest path visiting all vertices, starting and ending at the designated vertices. 
\end{definition}
For the sake of this article, the graph does not need to be complete. The distance between two non-adjacent vertices therefore is the length of the shortest path between them.  
In an online setting, where the graph is not known in advance but is given during the tour by revealing the neighborhood of each vertex, the problem is also known as \emph{Graph Exploration Problem}. 
Note that in an online setting, where not every information is revealed at the beginning, the shortest path between two points might not be clear in the beginning.

In this article, we focus on a variant where the structure of the graph is known in advance, but the actual edge weights only get revealed whenever an incident vertex is visited.  
\begin{definition}[Graph Exploration with Edge Weight Estimates (GEEWE)]
	Given a graph $G=(V,E)$, an upper and lower bound $u(e)$ and $\ell(e)$ for each edge weight $w(e)$, a vertex $s \in V$ as a start-vertex and a distinct vertex $t \in V$ as an end-vertex, the \emph{Graph Exploration with Edge Weight Estimates (GEEWE)} is the problem of finding the shortest walk visiting all vertices, starting and ending at the designated vertices. The actual weight $w(e) \in [\ell(e),u(e)]$ of an edge $e$ is revealed as soon as the salesman visits an incident vertex.
\end{definition}
While in the literature usually the start- and end-vertex are identical, we will assume that they are distinct from each other as in the travelling salesman path problem. For this sake, a \emph{walk} will refer to an open walk with distinct start and end vertices throughout the article. This helps avoiding minor technical issues like dealing with edge weights that got revealed but connect two vertices that both need to be visited (one of them being the starting vertex). The results however translate also to the classical travelling salesman problem, as two adjacent vertices can be chosen as start- and end vertex. The competitive ratio proven in the upper and lower bounds then stays almost the same, and only deviates by one potentially optimal chosen edge in the online algorithm's solution for all of the presented algorithms and lower bound constructions.

\subsection{Related Work}

\subsubsection{Travelling Salesman Problem.}
The problem of finding a shortest possible tour that visits a given set of places is basically as old as persons travel. The first explicit mathematical formulation is attributed to Menger in 1930 \cite{SCHRIJVER20051}, afterwards it became one of the most influential problems in discrete optimization. 

Due to its close relation to the Hamiltonian cycle problem, the travelling salesman problem is one of the first problems for which NP-hardness was known \cite{Karp1972}. 
As an early combinatorial optimization problem, it heavily influenced research in the area of integer programming, and was one of the first problems where the cutting plane method was proven to be valuable by Dantzig et al.~\cite{Dantzig54}.
On the approximation side, the Christofides-Serdyukov algorithm achieves a $1.5$-approximation ratio with a simple combinatorial argument in 1976 \cite{christofides}. For a long time this approximation ratio was the best possible for metric TSP, only recently the factor of $1.5$ from Christofides was improved to $1.5-\varepsilon$ by Karlin et al.~\cite{Karlin2021}. 
In practice, many real-world problems are solvable close to optimality.

Due to its natural setting, many variants of the travelling salesman problem are known and have real-life applications as well. This includes for example the vehicle routing problem, where more than just one person can traverse throughout the graph, or variants in which additional constraints for the visits of the salesperson at the vertices are added \cite{bellmore68,liu2023heuristicsvehicleroutingproblem}.

Due to its nature, many real-life applications in which the travelling salesman problem appears are in fact online settings. One track of the research for the online travelling salesman problem is the setting in which the actual requests are not known in advance, and only become available over time. If there are multiple salespersons, and requests need to be served immediately, this problem is also known as the \emph{k-server problem}, introduced by Manasse et al. \cite{manasse1988}. This problem focuses on which salesperson should serve the request. The question whether there is a $k$-competitive algorithm when $k$ salespersons are available is known as the \emph{$k$-server conjecture}, and is often referred to as the holy grail of online computation.

If however the requests do not have to be fulfilled immediately, the question about the order in which the request should be served becomes important. This line of research was started by Ausiello et al.~25 years ago \cite{ausiello01}. In this line of research the impact of predictions is studied as well, making the problem semi-online \cite{bampis23}. This recent article of Bampis et al. also gives a nice overview about this line of research.

The case where the structure of the graph is discovered online is called the \emph{graph exploration problem}.

\subsubsection{Graph Exploration.} 
The graph exploration setting usually refers to a problem, where an agent needs to visit all vertices of a given graph but is only aware of the graph induced by the so far visited vertices and its neighbors.

The study of this problem was initiated by Kalyanasundaram and Pruhs in 1994 \cite{KalyanasundaramConstructingCompetitiveTours1994}, where they proved a $16$-competitive ratio for planar graphs.

Since then, the graph exploration model attracted interest in various settings. It was shown by Miyazaki et al.~\cite{Miyazaki2009TheOG} that there cannot be an algorithm achieving a better competitive ratio than $2$ on unweighted graphs. The lower bound for the graph exploration problem on general graphs was then improved to $2.5$ by Dobrev et al.~\cite{DKM12}. 
Currently, the best lower bound is given by Birx et al.~\cite{BDHK21} with a competitive ratio of $10/3$ for planar graphs. 

On the other hand, no algorithm achieving a constant competitive ratio is known for general graphs. The so far best approach by Rosenkrantz et al.~\cite{Rosenkrantz1977} achieves a logarithmic competitive ratio, using a nearest neighbor approach. Another algorithm that achieves a logarithmic competitive ratio is a modification of depth-first search by Megow et al.~\cite{MegowOnlineGraphExploration2012a}. However, constant competitive algorithms are known for restricted graph classes.

In Megow et al.~\cite{MegowOnlineGraphExploration2012a}, an algorithm with constant competitive ratio for bounded genus graphs was presented. They also proved that this algorithm it is not constant competitive for general graphs. Thus, we may wonder whether knowing the structure of the graph in advance might help to construct exploration algorithms with constant competitive ratio. 

Their results are generalized by Baligacs et al.~\cite{BaligacsExplorationGraphsExcluded2023}, who prove that one can achieve constant competitive ratio on any minor-free graph class.

A recent study of the advice complexity of the unweighted case with unknown graph structure is provided in Böckenhauer et al.~\cite{BFU22}, where they show that an online algorithm can compute an optimal solution on sparse graphs if it is provided with $O(m)$ bits of correct advice, where $m$ is the number of edges.

In the same year, Eberle et al.~\cite{EberleRobustificationOnlineGraph2022a} studied the graph exploration problem using algorithms with predictions. Here, the authors present an algorithm that achieves a constant competitive ratio if the predictions are correct, but degrades the worse the predictions turn out to be. In real-world scenarios, these predictions could come from machine learning models trained on the kind of graphs that are expected.

Another research direction of the graph exploration problem is given by the setting where instead of one agent several can traverse through the graph simultaneously. Here, the goal is either to visit every vertex as soon as possible, or to minimize the total distance traversed by all agents. A recent overview about the different results in this direction can be found in van den Akker et al.~\cite{ABF24}.

\subsubsection{Online Algorithms with Estimates.}

In the research of online algorithms, one of the key questions is how to deal with the lack of information about the yet unknown parts of the problem. Therefore, a natural question that arises when thinking about an online problem is: ``What information is missing exactly?''

One could for example assume that basically nothing is known about the future input. This may be seen as the somewhat most intuitive point of view, indeed online problems were first introduced in this setting by Sleator and Tarjan \cite{SleatorT84}. For a general overview about online problems, we refer to the text books of Komm \cite{KommBook2016} or Borodin and El{-}Yaniv \cite{Borodin1998}.

Otherwise, one might assume that an algorithm does not have to be completely blind when attacking an online problem. Therefore, the concept of \emph{advice} was introduced by Böckenhauer et al.~\cite{Boeckenhauer2014}, where an algorithm may ask for some bits of arbitrary knowledge about the future input. This has lead to various theoretical insights, but also raised the question about the availability of such advice bits. One answer to this issue is the introduction of online algorithms with \emph{predictions} (sometimes called \emph{machine-learned advice}) by Purohit et al.~\cite{Purohit2018}, where the algorithm cannot ask for particular advice bits, but usually gets a hint on what to do with each part of the input. However, those hints may be wrong, and the algorithm's task is to balance the belief in the predictions but also keep some kind of backup if the hints turn out to be incorrect. 

This however assumes that the hint can be given in a local version. However, the prediction-giver will most likely deduce the hints of his belief from the global structure of the upcoming input - for example from some underlying machine learning program.

To model this setting, a variant of the prediction model was introduced that we will call online algorithms with \emph{estimates}. Here, a rough overview about the instance is given in advance, but the details like actual values of parts of the input are revealed in an online fashion.

A setting of this kind of estimates where parts of the instance are given but can be distorted at later points is natural for most problems that model a setting over time, for example scheduling. Here, the actual duration of each job might not be known at the beginning of those, like studied by Azar et al.~\cite{Azar2021,Azar2022-2,Azar2022} for different scheduling settings.

Another sensible setting that was modeled using our model of online algorithms with estimates are packing problems. Here, Gehnen et al.~\cite{Gehnen2024} studied the online knapsack problem, where an overview about the instance is given at the beginning, but each items size can deviate by a factor of $\alpha$. The actual weight of each item will be revealed when the algorithm needs to make a decision whether to take the item or not.

\section{Unrestricted Graph Classes}\label{chapter-general-case}
The graph exploration with edge weight estimates (GEEWE) problem is a generalization of the graph exploration problem. Therefore, lower bounds for the graph exploration problem also apply for the GEEWE problem, when allowing the range between upper and lower bound to be sufficiently large. 

On the other hand, every upper bound for the graph exploration problem is an upper bound for the GEEWE problem, no matter how small the ratio between upper and lower bound of each edge is. 

However, the lower bound constructions used for the graph exploration model also hide the structure of the graph, for example by presenting dead ends. As those structures can only be hidden in the GEEWE problem when the factor between upper and lower bound for some edges is very large, we focus on instances where this factor is rather small.

In this section, we will see that no algorithm can achieve a better competitive ratio than the largest factor $\alpha$ between the upper and lower edge weight ($u(e) \leq \alpha \ell (e)$ for all edges $e$), with $\alpha \leq 2$ . Furthermore, we show that this ratio can easily be achieved by a rather simple algorithm.

\newpage
\begin{theorem}\label{lower-bound-general}
	No algorithm for the GEEWE problem can achieve a better competitive ratio then the maximum factor $\alpha$ between an upper and lower bound of an announced edge weight when $\alpha \leq 2$.
\end{theorem}
\begin{proof}
	
	\textbf{Construction of the Instance.} 
	
	The following instance is constructed recursively, where a component of recursion depth $i$ consists of $k$ connected components of recursion depth $i-1$ for each $i \geq 1$ and a constant $k$. 
	
	A component of depth $0$ just consists of a path with $k+1$ vertices, where each edge is announced in the interval $[1,\alpha]$ and will get presented with the actual weight $1$. We will refer to the first vertex of the path as the starting vertex $s_0$, and to the last one as end vertex $t_0$.
	
	Each component of recursion depth $i \geq 1$ is constructed with two distinguished vertices as a start- and end vertex $s_i$ and $t_i$, and $k$ components of depth $i-1$ connected in a row between them. 
	The start vertex $s_i$ is connected to the vertices $s_{i-1}$ and $t_{i-1}$ of the first component of depth $i-1$ with an edge of announced weight $[k^i, \alpha k^i]$. Both will be revealed with the weight of $k^i$.
	
	Also, two adjacent components are connected with four edges of announced weight $[k^i, \alpha k^i]$, connecting both the start- and end vertex of one component to the start- and end vertex of the following component. The two edges connecting the start-vertex of the first visited component to the neighboring component will be revealed with weight $k^i$, the two edges incident to the end vertex have the actual weight $\alpha k^i$.
	
	Finally, the start- and end vertices of the last component of depth $i-1$ are connected to the end vertex $t_i$ with edges of announced weight $[k^i, \alpha k^i]$ as well. Again, the edge incident to the start vertex $s_{i-1}$ is revealed as $k^i$, the edge attached to the end vertex $t_{i-1}$ has actual size $\alpha k^i$. 
	
	We will call the edges that are announced with weights in the interval $[k^i, \alpha k^i]$ edges of level $i$. Note that this construction is symmetric, so based on the announcements the start- and end vertices of each component are not distinguishable from each other.

	Figure \ref{graph_lb_general} depicts an example for an instance of recursion depth $2$.

	\begin{figure}
	\centering
	\begin{tikzpicture}[scale=1.2]
	\draw 
	(1, 0) node[circle, black, draw](s2){$s_2$}
	(3, 2) node[circle, black, draw](s10){$s_1$}
	(3, -2) node[circle, black, draw](e10){$t_1$}
	(2, 1) node[circle, black, draw, scale=0.7](s000){$s_0$}
	(4, 1) node[circle, black, draw, scale=0.7](e000){$t_0$}
	(2.66,1)  node[circle, black,draw, scale=0.3](m001){$ $}
	(3.33,1)  node[circle, black, scale=0.3, draw](m002){$ $}
	(2, 0) node[circle, black, draw, scale=0.7](s010){$s_0$}
	(4, 0) node[circle, black, draw, scale=0.7](e010){$t_0$}
	(2.66,0)  node[circle, black, draw, scale=0.3](m011){$ $}
	(3.33,0)  node[circle, black, draw, scale=0.3](m012){$ $}
	(2, -1) node[circle, black, draw, scale=0.7](s020){$s_0$}
	(4, -1) node[circle, black, draw, scale=0.7](e020){$t_0$}
	(2.66,-1)  node[circle, black, draw, scale=0.3](m021){$ $}
	(3.33,-1)  node[circle, black, draw, scale=0.3](m022){$ $}
	(6, 2) node[circle, black, draw](s11){$s_1$}
	(6, -2) node[circle, black, draw](e11){$t_1$}
	(5, 1) node[circle, black, draw, scale=0.7](s030){$s_0$}
	(7, 1) node[circle, black, draw, scale=0.7](e030){$t_0$}
	(5.66,1)  node[circle, black,draw, scale=0.3](m031){$ $}
	(6.33,1)  node[circle, black, scale=0.3, draw](m032){$ $}
	(5, 0) node[circle, black, draw, scale=0.7](s040){$s_0$}
	(7, 0) node[circle, black, draw, scale=0.7](e040){$t_0$}
	(5.66,0)  node[circle, black, draw, scale=0.3](m041){$ $}
	(6.33,0)  node[circle, black, draw, scale=0.3](m042){$ $}
	(5, -1) node[circle, black, draw, scale=0.7](s050){$s_0$}
	(7, -1) node[circle, black, draw, scale=0.7](e050){$t_0$}
	(5.66,-1)  node[circle, black, draw, scale=0.3](m051){$ $}
	(6.33,-1)  node[circle, black, draw, scale=0.3](m052){$ $}
	(9, 2) node[circle, black, draw](s12){$s_1$}
	(9, -2) node[circle, black, draw](e12){$t_1$}
	(8, 1) node[circle, black, draw, scale=0.7](s060){$s_0$}
	(10, 1) node[circle, black, draw, scale=0.7](e060){$t_0$}
	(8.66,1)  node[circle, black,draw, scale=0.3](m061){$ $}
	(9.33,1)  node[circle, black, scale=0.3, draw](m062){$ $}
	(8, 0) node[circle, black, draw, scale=0.7](s070){$s_0$}
	(10, 0) node[circle, black, draw, scale=0.7](e070){$t_0$}
	(8.66,0)  node[circle, black, draw, scale=0.3](m071){$ $}
	(9.33,0)  node[circle, black, draw, scale=0.3](m072){$ $}
	(8, -1) node[circle, black, draw, scale=0.7](s080){$s_0$}
	(10, -1) node[circle, black, draw, scale=0.7](e080){$t_0$}
	(8.66,-1)  node[circle, black, draw, scale=0.3](m081){$ $}
	(9.33,-1)  node[circle, black, draw, scale=0.3](m082){$ $}

	(11, 0) node[circle, black, draw](e2){$t_2$};
	\draw[-] (s000) -- node[above, scale=0.6] {1} (m001);
	\draw[-] (m001) -- node[above, scale=0.6] {1} (m002);
	\draw[-] (m002) -- node[above, scale=0.6] {1} (e000);
	\draw[-] (s010) -- node[above, scale=0.6] {1} (m011);
	\draw[-] (m011) -- node[above, scale=0.6] {1} (m012);
	\draw[-] (m012) -- node[above, scale=0.6] {1} (e010);
	\draw[-] (s020) -- node[above, scale=0.6] {1} (m021);
	\draw[-] (m021) -- node[above, scale=0.6] {1} (m022);
	\draw[-] (m022) -- node[above, scale=0.6] {1} (e020);
	\draw[-] (s030) -- node[above, scale=0.6] {1} (m031);
	\draw[-] (m031) -- node[above, scale=0.6] {1} (m032);
	\draw[-] (m032) -- node[above, scale=0.6] {1} (e030);
	\draw[-] (s040) -- node[above, scale=0.6] {1} (m041);
	\draw[-] (m041) -- node[above, scale=0.6] {1} (m042);
	\draw[-] (m042) -- node[above, scale=0.6] {1} (e040);
	\draw[-] (s050) -- node[above, scale=0.6] {1} (m051);
	\draw[-] (m051) -- node[above, scale=0.6] {1} (m052);
	\draw[-] (m052) -- node[above, scale=0.6] {1} (e050);
	\draw[-] (s060) -- node[above, scale=0.6] {1} (m061);
	\draw[-] (m061) -- node[above, scale=0.6] {1} (m062);
	\draw[-] (m062) -- node[above, scale=0.6] {1} (e060);
	\draw[-] (s070) -- node[above, scale=0.6] {1} (m071);
	\draw[-] (m071) -- node[above, scale=0.6] {1} (m072);
	\draw[-] (m072) -- node[above, scale=0.6] {1} (e070);
	\draw[-] (s080) -- node[above, scale=0.6] {1} (m081);
	\draw[-] (m081) -- node[above, scale=0.6] {1} (m082);
	\draw[-] (m082) -- node[above, scale=0.6] {1} (e080);
	
	\draw[-] (s000) -- node[right, scale=0.6] {[3,6]} (s010);
	\draw[-] (e000) -- node[left, scale=0.6] {[3,6]} (e010);
	\draw[-] (s000) -- node[above, scale=0.6] {[3,6]} (e010);
	\draw[-] (e000) -- node[below, scale=0.6] {[3,6]} (s010);	
	\draw[-] (s010) -- node[right, scale=0.6] {[3,6]} (s020);
	\draw[-] (e010) -- node[left, scale=0.6] {[3,6]} (e020);
	\draw[-] (s010) -- node[above, scale=0.6] {[3,6]} (e020);
	\draw[-] (e010) -- node[below, scale=0.6] {[3,6]} (s020);
	
	\draw[-] (s030) -- node[right, scale=0.6] {[3,6]} (s040);
	\draw[-] (e030) -- node[left, scale=0.6] {[3,6]} (e040);
	\draw[-] (s030) -- node[above, scale=0.6] {[3,6]} (e040);
	\draw[-] (e030) -- node[below, scale=0.6] {[3,6]} (s040);
	\draw[-] (s040) -- node[right, scale=0.6] {[3,6]} (s050);
	\draw[-] (e040) -- node[left, scale=0.6] {[3,6]} (e050);
	\draw[-] (s040) -- node[above, scale=0.6] {[3,6]} (e050);
	\draw[-] (e040) -- node[below, scale=0.6] {[3,6]} (s050);
	
	\draw[-] (s060) -- node[right, scale=0.6] {[3,6]} (s070);
	\draw[-] (e060) -- node[left, scale=0.6] {[3,6]} (e070);
	\draw[-] (s060) -- node[above, scale=0.6] {[3,6]} (e070);
	\draw[-] (e060) -- node[below, scale=0.6] {[3,6]} (s070);
	\draw[-] (s070) -- node[right, scale=0.6] {[3,6]} (s080);
	\draw[-] (e070) -- node[left, scale=0.6] {[3,6]} (e080);
	\draw[-] (s070) -- node[above, scale=0.6] {[3,6]} (e080);
	\draw[-] (e070) -- node[below, scale=0.6] {[3,6]} (s080);
	\draw[-] (s10) -- node[right, scale=0.6] {[3,6]} (s000);
	\draw[-] (s10) -- node[left, scale=0.6] {[3,6]} (e000);
	\draw[-] (s11) -- node[right, scale=0.6] {[3,6]} (s030);
	\draw[-] (s11) -- node[left, scale=0.6] {[3,6]} (e030);
	\draw[-] (s12) -- node[right, scale=0.6] {[3,6]} (s060);
	\draw[-] (s12) -- node[left, scale=0.6] {[3,6]} (e060);
	\draw[-] (s020) -- node[right, scale=0.6] {[3,6]} (e10);
	\draw[-] (e020) -- node[left, scale=0.6] {[3,6]} (e10);
	\draw[-] (s050) -- node[right, scale=0.6] {[3,6]} (e11);
	\draw[-] (e050) -- node[left, scale=0.6] {[3,6]} (e11);	
	\draw[-] (s080) -- node[right, scale=0.6] {[3,6]} (e12);
	\draw[-] (e080) -- node[left, scale=0.6] {[3,6]} (e12);
	\draw[-] (s2) ..controls (1,2)  ..  (s10);
	\draw[-] (s2) ..controls (1,-2)  ..  (e10);
	\draw[-] (e2) ..controls (11,2)  ..  (s12);
	\draw[-] (e2) ..controls (11,-2)  ..  (e12);
	\draw[-] (s10) ..controls (5.4,1)  and  (3.6,-1) .. (e11);
	\draw[-] (s11) ..controls (8.4,1)  and  (6.6,-1) .. (e12);
	\draw[-] (e10) ..controls (5.4,-1)  and  (3.6,1) .. (s11);
	\draw[-] (e11) ..controls (8.4,-1)  and  (6.6,1) .. (s12);
	\draw[-] (s10) -- node[above, scale=0.6] {[9,18]} (s11);
	\draw[-] (s11) -- node[above, scale=0.6] {[9,18]} (s12);
	\draw[-] (e10) -- node[below, scale=0.6] {[9,18]} (e11);
	\draw[-] (e11) -- node[below, scale=0.6] {[9,18]} (e12);
	\draw
	(1, 2) node[scale=0.6](n1){[9,18]}
	(1, -2) node[scale=0.6](n2){[9,18]}
	(11, 2) node[scale=0.6](n3){[9,18]}
	(11, -2) node[scale=0.6](n4){[9,18]}
	(4.2, 1.7) node[scale=0.6](n5){[9,18]}
	(4.2, -1.7) node[scale=0.6](n5){[9,18]}
	(7.2, 1.7) node[scale=0.6](n5){[9,18]}
	(7.2, -1.7) node[scale=0.6](n5){[9,18]};
	
	\end{tikzpicture}
	
		\caption{An example instance with the announcements of recursion depth $2$ with $k=3$ and $\alpha=2$. An online algorithm needs to traverse from $s_2$ to $t_2$. The edges of level $0$ are just given with the weight $1$, as they will be presented with weight $1$ anyway. Depending on the decisions of the algorithm, the other edges will either have a value of the upper or lower bound.}\label{graph_lb_general}
	\end{figure}
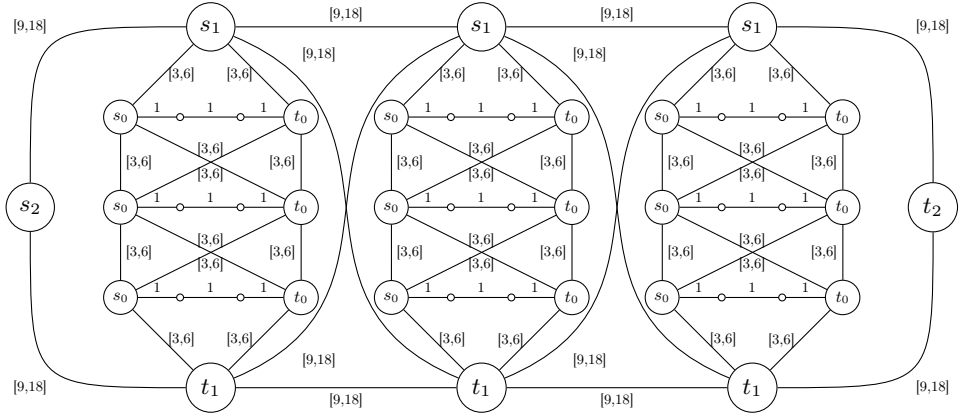
	As the start- and end vertices of each component are not distinguishable to the algorithm by the announcements, we can say without loss of generality that the algorithm reaches the start vertex of each component before reaching the end vertex.
	
	\textbf{How an algorithm traverse the Instance.} 
	
	We claim that no online algorithm can traverse from the start to the end vertex of a component of recursion depth $i\geq 0$ visiting each vertex within with costs of less than $i (\alpha k +1)k^i + k^{i+1}$, while an optimal offline solution is able to do the traversal with total costs of $i (k +1)k^i + k^{i+1}$.
	
	For level $i=0$, it is easy to see that both the online and offline algorithm can traverse through the path, resulting in total costs of $k$.
	
	For each level $i \geq 1$, the algorithm needs to traverse through $k$ components of recursion depth $i-1$, and needs to take at least $k+1$ edges from the $i$-th level. As the $k+1$ connections between two consecutive components cost at least $k^i$ each, and traversing each of the $k$ components costs at least $(i-1) (\alpha k +1)k^{i-1} + k^{i}$ by the induction hypothesis, the algorithm needs to pay at least $ k ( (i-1) (\alpha k +1)k^{i-1} + k^{i} ) + (k+1) k^i = (k+1) k^i + (i-1) (\alpha k +1)k^i + k^{i+1} $. However, this calculation only works if the algorithm traverses between the components on the cheap edges with costs of $k^i$ and traverses inside the components just the necessary amount.
	
	To reach a new component, the algorithm can either leave the previous component at the end vertex with costs of $\alpha k^i$, or leave the previous component at its start vertex with costs of $k^i$. In the former case, this traversal costs $(\alpha-1)k^i$ more than in the previous calculation.
	 
	In the latter case, the algorithm still needs to visit all the vertices in the previous component, including reaching its end vertex. This can be done by traversing in this component to the end vertex and back to the start vertex, or by visiting the component at least twice. If the algorithm traverses inside the component to the end and back, it needs to pay for the traversal back at least the costs of an offline algorithm, so $(i-1) (k +1)k^{i-1} + k^{i}$ in addition to the previous calculation.
	
	If the algorithm finally decides to visit the component twice, additional cost of a cheap edge of level $i$ with costs of $k^i$ are unavoidable for an additional movement between components. 
	
	Together, the algorithm must pay additional costs of at least $(\alpha-1)k^i$ for each component, resulting in total costs of at least $k (\alpha-1)k^i + (k+1) k^i + (i-1) (\alpha k +1)k^i + k^{i+1} = i (\alpha k +1)k^i + k^{i+1}$. 
	
	An optimal solution however arrives at the end vertex of each component first, thus being able to leave each component of recursion depth $i$ with the cheap $k^i$ edge from its start vertex to the end vertex of the next component. Overall, a traversal of a depth $i$ component costs $k+1$ times $k^i$ for the level $i$ traversals, and $k$ traversal of the depth $i-1$ components. This sums up to $ k ( (i-1) (k +1)k^{i-1} + k^{i}) + (k+1) k^i = i (k +1)k^i + k^{i+1} $. 
		
	Overall, for a given $k$, no online algorithm is able to achieve a better competitive ratio than $\frac{i (\alpha k +1)k^i + k^{i+1}}{i (k +1)k^i + k^{i+1}} = \frac{i (\alpha k +1)+k}{i (k+1)+k}$ to traverse through a component of recursion depth $i$. It shows that there cannot be a constant competitive ratio of less than $\alpha$, when $i$ and $k$ are chosen sufficiently large.
	\qed
\end{proof}
\begin{remark}
	Remark that if $\alpha \geq 2$, we still get a lower bound of $2$ by applying the proof of Theorem~\ref{lower-bound-general}.
\end{remark}
This competitive ratio can be achieved by a trivial algorithm that completely ignores the revealed edge weights:

\begin{theorem}\label{th:alpha}
	An algorithm that precomputes the shortest possible walk covering all vertices from start to end vertex, based on the lower bounds of the edges and sticks to it, is at most $\alpha$ times worse than an optimal tour when $\alpha$ is the maximum factor between the announced bounds of an edge. 
\end{theorem} 
\begin{proof}
	The precomputed walk can be at most $\alpha$ times more expensive than the sum of the lower bounds of the edges. However, since no solution can be better than the precomputed best tour based on the lower bounds, the precomputed walk is at most $\alpha$ times more expensive than any valid travelling salesman walk.
	\qed
\end{proof}

Note that the running time of this algorithm will probably be exponential in the worst case, as it needs to solve an NP-hard problem. 

\section{Uniform announced Edge Weights and restricted Graph Classes}\label{chapter-special-case}

As we have seen in the previous section, we cannot hope for an algorithm that beats the (from the online perspective) trivial precomputing algorithm in general. Therefore, it is interesting to see if there are algorithms that actually perform better on restricted graph classes, using the revealed edge weights during the algorithm's traversal throughout the graph.
A common restriction we will use in this section is the assumption that the graph has uniform announced edge weights $[1,\alpha]$ for all edges. This allows us to achieve improved results e.g.~for complete graphs, as without restrictions of the edge weights every graph structure can implicitly be constructed by using edges of arbitrary large announced weight.
\newpage
\subsection{Recalculating Algorithm}

An intuitive candidate is the following algorithm, that recalculates the cheapest possible walk visiting all vertices, at each step taking the newly revealed edges into account.

\begin{algorithm}
	 \floatname{algorithm}{Adaptive Exploration Algorithm}
	\caption{}
	\begin{algorithmic}
		\Require start vertex $s$, end vertex $t$
		\State Agent $A$ starts at vertex $s$
		\While{Graph is not fully explored}
		\State Calculate a worst case shortest walk from $A$ to $t$, visiting all unvisited vertices.
		\State Move $A$ to the next vertex on the walk.  
		\EndWhile
	\end{algorithmic}
\end{algorithm}
 In every step, the algorithm can calculate an upper bound for finishing the walk by using the revealed edge weights together with the upper bound of each weight for the unrevealed edge weights. Calculating the walk is computationally expensive, as the algorithm needs to recalculate the optimal for each visited vertex.
 
 This algorithm can be proven to be optimal for graph classes like complete graphs or complete balanced bipartite graphs.

\begin{theorem}\label{th:cr-complete-graph}
	Given a graph where each path can be extended to a Hamiltonian path from the start to a distinct end vertex and uniform announced edge weights $[1, \alpha]$ for $\alpha < 2$, the adaptive exploration algorithm reaches a competitive ratio of at most $\frac{\alpha+1}{2}$.
\end{theorem}
We start with an easy observation, connecting the adaptive exploration algorithm with the nearest neighbor method.
\begin{lemma}\label{lemma6}
	In the setting of Theorem~\ref{th:cr-complete-graph}, the adaptive exploration algorithm takes the cheapest outgoing edge towards an unvisited vertex (which is not the end vertex), until it visits the end vertex in its final step.
\end{lemma}
Note that in general the nearest neighbor method \cite{Rosenkrantz1977} is not identical with the adaptive exploration algorithm, as the nearest neighbor method does not take the global graph structure into account. For example, as the adaptive exploration algorithm computes a walk based on the worst case announcements, it will achieve a competitive ratio of at most $\alpha$, while the nearest neighbor method might only achieve a logarithmic approximation.

\begin{proof}[of Lemma~\ref{lemma6}]
	At any point there is a Hamiltonian extension to the current solution. Therefore, at any point the algorithm has an option with expected best case costs of the next edge plus the amount of remaining vertices minus one (of which the weight to reach them is unknown between $1$ and $\alpha$). This beats every option that involves visiting vertices twice, so no vertex will be visited twice by the adaptive exploration algorithm.
	
	As all outgoing edges towards unvisited vertices (that are not the end vertex) are part of Hamiltonian extensions whose expected worst case costs to the algorithm only deviate by its first edge, the decision which vertex is visited next is determined by the cheapest outgoing edge.
	\qed
\end{proof}

Given this observed behaviour of the adaptive exploration algorithm, we are able to prove \Cref{th:cr-complete-graph}: 

\begin{proof}[of \Cref{th:cr-complete-graph}]
	First, we fix an optimal walk for the analysis.
	
	For the sake of notation, we enumerate the vertices from $1$ to $n$ in the same order as they are visited by the adaptive exploration algorithm. To prove the claim, assume that after $k$ vertices the algorithm has paid the edge costs of $w_1, \dots , w_{k-1}$. The optimal solution must be at least twice the cheaper half of the used edges of the algorithm:
	
	Since the adaptive exploration algorithm takes the cheapest possible outgoing edge towards an unvisited vertex, we can assume that every edge leaving a vertex $x$ towards a vertex $y$ with $y > x$ costs at least $w_x$.
	Edges to already visited vertices can be cheaper. As the optimal walk is a Hamiltonian one, only two edges incident to each vertex can be used in an optimal solution.
	
	When comparing a fixed optimal solution to the walk of the adaptive exploration algorithm, there are two types of edges this optimal solution consists of: Edges from a vertex $x$ to a vertex $y$ with $x<y$, and edges from $x$ to $y$ with $x>y$.
	
	For each edge with $x<y$, by Lemma~\ref{lemma6}, we know that the algorithm has paid at most as much as the optimal solution when leaving $x$. 
	
	For each edge with $x>y$, it follows by Lemma~\ref{lemma6}, that the algorithm has paid at most as much when leaving $y$ as the optimal solution paid for leaving $x$. 
	
	Sadly, both cannot be combined as edges might have counted in both cases. But as the optimal solution needs $n-1$ edges in total, one of the two types of edges must occur at least $\frac{n-1}{2}$ times. Therefore, half the edges of the algorithm's solution cost as much as half the edges of the optimal solution. Let's say those edge weights total to $K$.
	
	The costs of an optimal solution are now at least $\frac{n-1}{2}+K$, while the algorithm's solution is at most $\frac{\alpha (n-1)}{2}+K$. The ratio between those expressions is largest possible when $K$ is small. However, $K$ must be at least $\frac{n-1}{2}$.
	It follows that the competitive ratio of the adaptive exploration algorithm is at most $\frac{\alpha+1}{2}$.
	\qed
\end{proof}

Even if the restriction for the graph classes is rather strict, there are some important graph classes where \Cref{th:cr-complete-graph} applies.
\begin{corollary}
	For every $\alpha < 2$ and every $n$, the adaptive exploration algorithm achieves a competitive ratio of $\frac{\alpha+1}{2}$ on the complete graphs $K_n$ with uniform announced edge weight and distinct start- and end vertices.
	\end{corollary}
\newpage
\begin{corollary}\label{algbpgraph}
	For every $\alpha < 2$ and every $n$, the adaptive exploration algorithm achieves a competitive ratio of $\frac{\alpha+1}{2}$ on the complete bipartite graphs $K_{n,n}$ with uniform announced edge weights, when start- and end vertices are on different sides, and on the complete bipartite graph $K_{n+1,n}$ with distinct start- and end vertices on the bigger side.
\end{corollary}
Both corollaries immediately follow from \Cref{th:cr-complete-graph}. 

Indeed, the performance of the adaptive exploration algorithm is optimal on those graph classes.

\begin{theorem}\label{lower-bound-complete}
	No algorithm can achieve a better competitive ratio than $\frac{\alpha+1}{2}$ even on complete graphs with uniform announced edge weights $[1,\alpha]$ and $\alpha \leq 2$.
\end{theorem}

\begin{proof}
	Let us consider any algorithm $\mathcal{A}$ on the complete graph $K_{2k}$. For the first $k$ rounds, we reveal that every edge adjacent to the current vertex has a weight of $1$. We call the set of vertices visited in this phase $A$. Then, every newly revealed edge has a weight $\alpha$ (the edges going to vertices in $A$ are still of weight $1$). We set $B=V\setminus A$ the set of vertices visited in this second phase.
	
	Here, we use weights $1$ and $\alpha$, so it never makes sense for an algorithm to go back to an already visited edge, because the triangle inequality is satisfied. Indeed, imagine that $\mathcal{A}$ follows the path $v_1\cdots v_k$ (for $k\geq 3$) at some point but had already visited $v_2\cdots v_{k-1}$ when it was at $v_1$. Then, the cost of going from $v_1$ to $v_k$ directly is at most $\alpha$, whereas that of the path is at least $(k-1)\geq 2\geq \alpha$. Thus, we can assume that $|A|=|B|=k$.
	
	The cost of the path computed by the algorithm is $(1+\alpha)k$, whereas there exists an optimal path of cost $2k$: with $A=\{a_1, \cdots, a_k\}, B=\{b_1, \cdots, b_k\}$, the path $a_1b_1a_2b_2\cdots a_kb_ka_1$ only takes edges adjacent to vertices in $A$, and which are thus of weight $1$. Thus, the algorithm cannot do better than a competitive ratio of $\frac{1+\alpha}2$.
	\qed
\end{proof}

\begin{remark}
	Remark that if $\alpha > 2$, we still get a lower bound of $3/2$ by applying the proof of \Cref{lower-bound-complete}.
\end{remark}
A lower bound of $\frac{\alpha+1}{2}$ for complete bipartite graphs $K_{n,n}$, matching \Cref{algbpgraph}, can be constructed analogously.
Note that the restriction for distinct start- and end vertices is only necessary for achieving a strict bound of $\frac{1+\alpha}2$, in case of identical start- and end vertex the ratio can decrease by $\frac{\alpha-1}{n}$ as the final edge leading towards the start/end vertex is revealed as of the beginning, so the algorithm cannot be tricked into taking the last edge as an expensive edge.

However, the adaptive exploration algorithm is not able to outperform the trivial precomputing algorithm even for uniform announced edge weights and $\alpha < 2$, even on grids.
\newpage
\begin{theorem}
	The Adaptive Exploration Algorithm does not achieve a better competitive ratio than $\alpha$ for graphs with announced weight $[1,\alpha]$ for all edges and $\alpha < 2$ in general.
\end{theorem}

\begin{proof} 
We can construct a grid as an example: 
\tikzset{opt/.style={line width=0.1mm}} 
\tikzset{optalg/.style={line width=1mm}} 
\tikzset{alg/.style={dashed, line width=1mm}} 
\tikzset{unused/.style={dashed, line width=0.1mm}} 
\tikzset{showit/.style={decorate, decoration=triangles}}
	\begin{figure}
	\centering
	\begin{tikzpicture}[scale=0.80]
	\draw 
	(1, 0) node[circle, black, draw](00){}
	(3, 0) node[circle, black, draw](01){}
	(5, 0) node[circle, black, draw](02){}
	(7, 0) node[circle, black, draw](03){}
	(9, 0) node[circle, black, draw](04){}
	(11, 0) node[circle, black, draw](05){}
	(13, 0) node[circle, black, draw](06){}
	
	(1, -2) node[circle, black, draw](20){} 
	(3, -2) node[circle, black, draw](21){}
	(5, -2) node[circle, black, draw](22){}
	(7, -2) node[circle, black, draw](23){}
	(9, -2) node[circle, black, draw](24){}
	(11,-2) node[circle, black, draw](25){}
	(13,-2) node[circle, black, draw](26){}
		(15,-2) node[circle, black, draw](27){}
	
	(-1, -4) node[circle, black, draw](4m1){}
	(1, -4) node[circle, black, draw](40){}
	(3, -4) node[circle, black, draw](41){}
	(5, -4) node[circle, black, draw](42){}
	(7, -4) node[circle, black, draw](43){}
	(9, -4) node[circle, black, draw](44){}
	(11,-4) node[circle, black, draw](45){}
	(13,-4) node[circle, black, draw](46){}
	
	(1, -6) node[circle, black, draw](60){}
	(3, -6) node[circle, black, draw](61){}
	(5, -6) node[circle, black, draw](62){}
	(7, -6) node[circle, black, draw](63){}
	(9, -6) node[circle, black, draw](64){}
	(11,-6) node[circle, black, draw](65){}
	(13,-6) node[circle, black, draw](66){}
	(15,-6) node[circle, black, draw](67){}
	
	(-1, -8) node[circle, black, draw](8m1){}
	(1, -8) node[circle, black, draw](80){}
	(3, -8) node[circle, black, draw](81){}
	(5, -8) node[circle, black, draw](82){}
	(7, -8) node[circle, black, draw](83){}
	(9, -8) node[circle, black, draw](84){}
	(11,-8) node[circle, black, draw](85){}
	(13,-8) node[circle, black, draw](86){}
	
	(0, -1) node[circle, black, draw](10){{$s$}} 
	(2, -1) node[circle, black, draw](11){}
	(4, -1) node[circle, black, draw](12){}
	(6, -1) node[circle, black, draw](13){}
	(8, -1) node[circle, black, draw](14){}
	(10,-1) node[circle, black, draw](15){}
	(12,-1) node[circle, black, draw](16){}
	(14,-1) node[circle, black, draw](17){}
	
	(0, -3) node[circle, black, draw](30){}
	(2, -3) node[circle, black, draw](31){}
	(4, -3) node[circle, black, draw](32){}
	(6, -3) node[circle, black, draw](33){}
	(8, -3) node[circle, black, draw](34){}
	(10,-3) node[circle, black, draw](35){}
	(12,-3) node[circle, black, draw](36){}
	(14,-3) node[circle, black, draw](37){}	
	
	(0, -5) node[circle, black, draw](50){}
	(2, -5) node[circle, black, draw](51){}
	(4, -5) node[circle, black, draw](52){}
	(6, -5) node[circle, black, draw](53){}
	(8, -5) node[circle, black, draw](54){}
	(10,-5) node[circle, black, draw](55){}
	(12,-5) node[circle, black, draw](56){}
	(14,-5) node[circle, black, draw](57){}
		
	(0, -7) node[circle, black, draw](70){}
	(2, -7) node[circle, black, draw](71){}
	(4, -7) node[circle, black, draw](72){}
	(6, -7) node[circle, black, draw](73){}
	(8, -7) node[circle, black, draw](74){}
	(10,-7) node[circle, black, draw](75){}
	(12,-7) node[circle, black, draw](76){}
	(14,-7) node[circle, black, draw](77){}
	
	(0, -9) node[circle, black, draw](90){}
	(2, -9) node[circle, black, draw](91){}
	(4, -9) node[circle, black, draw](92){}
	(6, -9) node[circle, black, draw](93){}
	(8, -9) node[circle, black, draw](94){}
	(10,-9) node[circle, black, draw](95){}
	(12,-9) node[circle, black, draw](96){}
	(14,-9) node[circle, black, draw](97){$t$};
	\draw[optalg] (10) -- node[left,scale=0.8, black] {$\alpha$} (00);
	\draw[optalg] (11) -- node[left,scale=0.8, black] {$\alpha$} (01);
	\draw[optalg] (12) -- node[left,scale=0.8, black] {$\alpha$} (02);
	\draw[optalg] (13) -- node[left,scale=0.8, black] {$\alpha$} (03);
	\draw[optalg] (14) -- node[left,scale=0.8, black] {$\alpha$} (04);
	\draw[optalg] (15) -- node[left,scale=0.8, black] {$\alpha$} (05);
	\draw[optalg] (16) -- node[left,scale=0.8, black] {$\alpha$} (06);
	\draw[optalg] (11) -- node[left,scale=0.8, black] {$\alpha$} (00);
	\draw[optalg] (12) -- node[left,scale=0.8, black] {$\alpha$} (01);
	\draw[optalg] (13) -- node[left,scale=0.8, black] {$\alpha$} (02);
	\draw[optalg] (14) -- node[left,scale=0.8, black] {$\alpha$} (03);
	\draw[optalg] (15) -- node[left,scale=0.8, black] {$\alpha$} (04);
	\draw[optalg] (16) -- node[left,scale=0.8, black] {$\alpha$} (05);
	\draw[optalg] (17) -- node[left,scale=0.8, black] {$\alpha$} (06);
	
	\draw[optalg] (17) -- node[left,scale=0.8, black] {$\alpha$} (27);
	\draw[optalg] (37) -- node[left,scale=0.8, black] {$\alpha$} (27);
	
	\draw[optalg] (30) -- node[left,scale=0.8, black] {$\alpha$} (4m1);
	\draw[optalg] (50) -- node[left,scale=0.8, black] {$\alpha$} (4m1);
	
	\draw[optalg] (57) -- node[left,scale=0.8, black] {$\alpha$} (67);
	\draw[optalg] (77) -- node[left,scale=0.8, black] {$\alpha$} (67);
	
	\draw[optalg] (70) -- node[left,scale=0.8, black] {$\alpha$} (8m1);
	\draw[optalg] (90) -- node[left,scale=0.8, black] {$\alpha$} (8m1);
	
	\draw[unused] (10) -- node[left,scale=0.8, black] {1} (20);
	\draw[opt] (11) -- node[left,scale=0.8, black] {1} (21);
	\draw[opt] (12) -- node[left,scale=0.8, black] {1} (22);
	\draw[opt] (13) -- node[left,scale=0.8, black] {1} (23);
	\draw[opt] (14) -- node[left,scale=0.8, black] {1} (24);
	\draw[opt] (15) -- node[left,scale=0.8, black] {1} (25);
	\draw[opt] (16) -- node[left,scale=0.8, black] {1} (26);
	\draw[opt] (11) -- node[left,scale=0.8, black] {1} (20);
	\draw[opt] (12) -- node[left,scale=0.8, black] {1} (21);
	\draw[opt] (13) -- node[left,scale=0.8, black] {1} (22);
	\draw[opt] (14) -- node[left,scale=0.8, black] {1} (23);
	\draw[opt] (15) -- node[left,scale=0.8, black] {1} (24);
	\draw[opt] (16) -- node[left,scale=0.8, black] {1} (25);
	\draw[unused] (17) -- node[left,scale=0.8, black] {1} (26);
	
	\draw[optalg] (30) -- node[left,scale=0.8, black] {$\alpha$} (20);
	\draw[alg] (31) -- node[left,scale=0.8, black] {$\alpha$} (21);
	\draw[alg] (32) -- node[left,scale=0.8, black] {$\alpha$} (22);
	\draw[alg] (33) -- node[left,scale=0.8, black] {$\alpha$} (23);
	\draw[alg] (34) -- node[left,scale=0.8, black] {$\alpha$} (24);
	\draw[alg] (35) -- node[left,scale=0.8, black] {$\alpha$} (25);
	\draw[alg] (36) -- node[left,scale=0.8, black] {$\alpha$} (26);
	\draw[alg] (31) -- node[left,scale=0.8, black] {$\alpha$} (20);
	\draw[alg] (32) -- node[left,scale=0.8, black] {$\alpha$} (21);
	\draw[alg] (33) -- node[left,scale=0.8, black] {$\alpha$} (22);
	\draw[alg] (34) -- node[left,scale=0.8, black] {$\alpha$} (23);
	\draw[alg] (35) -- node[left,scale=0.8, black] {$\alpha$} (24);
	\draw[alg] (36) -- node[left,scale=0.8, black] {$\alpha$} (25);
	\draw[optalg] (37) -- node[left,scale=0.8, black] {$\alpha$} (26);
	
	\draw[unused] (30) -- node[left,scale=0.8, black] {1} (40);
	\draw[opt] (31) -- node[left,scale=0.8, black] {1} (41);
	\draw[opt] (32) -- node[left,scale=0.8, black] {1} (42);
	\draw[opt] (33) -- node[left,scale=0.8, black] {1} (43);
	\draw[opt] (34) -- node[left,scale=0.8, black] {1} (44);
	\draw[opt] (35) -- node[left,scale=0.8, black] {1} (45);
	\draw[opt] (36) -- node[left,scale=0.8, black] {1} (46);
	\draw[opt] (31) -- node[left,scale=0.8, black] {1} (40);
	\draw[opt] (32) -- node[left,scale=0.8, black] {1} (41);
	\draw[opt] (33) -- node[left,scale=0.8, black] {1} (42);
	\draw[opt] (34) -- node[left,scale=0.8, black] {1} (43);
	\draw[opt] (35) -- node[left,scale=0.8, black] {1} (44);
	\draw[opt] (36) -- node[left,scale=0.8, black] {1} (45);
	\draw[unused] (37) -- node[left,scale=0.8, black] {1} (46);
	
	\draw[optalg] (50) -- node[left,scale=0.8, black] {$\alpha$} (40);
	\draw[alg] (51) -- node[left,scale=0.8, black] {$\alpha$} (41);
	\draw[alg] (52) -- node[left,scale=0.8, black] {$\alpha$} (42);
	\draw[alg] (53) -- node[left,scale=0.8, black] {$\alpha$} (43);
	\draw[alg] (54) -- node[left,scale=0.8, black] {$\alpha$} (44);
	\draw[alg] (55) -- node[left,scale=0.8, black] {$\alpha$} (45);
	\draw[alg] (56) -- node[left,scale=0.8, black] {$\alpha$} (46);
	\draw[alg] (51) -- node[left,scale=0.8, black] {$\alpha$} (40);
	\draw[alg] (52) -- node[left,scale=0.8, black] {$\alpha$} (41);
	\draw[alg] (53) -- node[left,scale=0.8, black] {$\alpha$} (42);
	\draw[alg] (54) -- node[left,scale=0.8, black] {$\alpha$} (43);
	\draw[alg] (55) -- node[left,scale=0.8, black] {$\alpha$} (44);
	\draw[alg] (56) -- node[left,scale=0.8, black] {$\alpha$} (45);
	\draw[optalg] (57) -- node[left,scale=0.8, black] {$\alpha$} (46);
	\draw[unused] (50) -- node[left,scale=0.8, black] {1} (60);
	\draw[opt] (51) -- node[left,scale=0.8, black] {1} (61);
	\draw[opt] (52) -- node[left,scale=0.8, black] {1} (62);
	\draw[opt] (53) -- node[left,scale=0.8, black] {1} (63);
	\draw[opt] (54) -- node[left,scale=0.8, black] {1} (64);
	\draw[opt] (55) -- node[left,scale=0.8, black] {1} (65);
	\draw[opt] (56) -- node[left,scale=0.8, black] {1} (66);
	\draw[opt] (51) -- node[left,scale=0.8, black] {1} (60);
	\draw[opt] (52) -- node[left,scale=0.8, black] {1} (61);
	\draw[opt] (53) -- node[left,scale=0.8, black] {1} (62);
	\draw[opt] (54) -- node[left,scale=0.8, black] {1} (63);
	\draw[opt] (55) -- node[left,scale=0.8, black] {1} (64);
	\draw[opt] (56) -- node[left,scale=0.8, black] {1} (65);
	\draw[unused] (57) -- node[left,scale=0.8, black] {1} (66);
	
	\draw[optalg] (70) -- node[left,scale=0.8, black] {$\alpha$} (60);
	\draw[alg] (71) -- node[left,scale=0.8, black] {$\alpha$} (61);
	\draw[alg] (72) -- node[left,scale=0.8, black] {$\alpha$} (62);
	\draw[alg] (73) -- node[left,scale=0.8, black] {$\alpha$} (63);
	\draw[alg] (74) -- node[left,scale=0.8, black] {$\alpha$} (64);
	\draw[alg] (75) -- node[left,scale=0.8, black] {$\alpha$} (65);
	\draw[alg] (76) -- node[left,scale=0.8, black] {$\alpha$} (66);
	\draw[alg] (71) -- node[left,scale=0.8, black] {$\alpha$} (60);
	\draw[alg] (72) -- node[left,scale=0.8, black] {$\alpha$} (61);
	\draw[alg] (73) -- node[left,scale=0.8, black] {$\alpha$} (62);
	\draw[alg] (74) -- node[left,scale=0.8, black] {$\alpha$} (63);
	\draw[alg] (75) -- node[left,scale=0.8, black] {$\alpha$} (64);
	\draw[alg] (76) -- node[left,scale=0.8, black] {$\alpha$} (65);
	\draw[optalg] (77) -- node[left,scale=0.8, black] {$\alpha$} (66);
	
	\draw[unused] (70) -- node[left,scale=0.8, black] {1} (80);
	\draw[opt] (71) -- node[left,scale=0.8, black] {1} (81);
	\draw[opt] (72) -- node[left,scale=0.8, black] {1} (82);
	\draw[opt] (73) -- node[left,scale=0.8, black] {1} (83);
	\draw[opt] (74) -- node[left,scale=0.8, black] {1} (84);
	\draw[opt] (75) -- node[left,scale=0.8, black] {1} (85);
	\draw[opt] (76) -- node[left,scale=0.8, black] {1} (86);
	\draw[opt] (71) -- node[left,scale=0.8, black] {1} (80);
	\draw[opt] (72) -- node[left,scale=0.8, black] {1} (81);
	\draw[opt] (73) -- node[left,scale=0.8, black] {1} (82);
	\draw[opt] (74) -- node[left,scale=0.8, black] {1} (83);
	\draw[opt] (75) -- node[left,scale=0.8, black] {1} (84);
	\draw[opt] (76) -- node[left,scale=0.8, black] {1} (85);
	\draw[unused] (77) -- node[left,scale=0.8, black] {1} (86);
	
	\draw[optalg] (90) -- node[left,scale=0.8, black] {$\alpha$} (80);
	\draw[optalg] (91) -- node[left,scale=0.8, black] {$\alpha$} (81);
	\draw[optalg] (92) -- node[left,scale=0.8, black] {$\alpha$} (82);
	\draw[optalg] (93) -- node[left,scale=0.8, black] {$\alpha$} (83);
	\draw[optalg] (94) -- node[left,scale=0.8, black] {$\alpha$} (84);
	\draw[optalg] (95) -- node[left,scale=0.8, black] {$\alpha$} (85);
	\draw[optalg] (96) -- node[left,scale=0.8, black] {$\alpha$} (86);
	\draw[alg] (91) -- node[left,scale=0.8, black] {$\alpha$} (80);
	\draw[alg] (92) -- node[left,scale=0.8, black] {$\alpha$} (81);
	\draw[alg] (93) -- node[left,scale=0.8, black] {$\alpha$} (82);
	\draw[alg] (94) -- node[left,scale=0.8, black] {$\alpha$} (83);
	\draw[alg] (95) -- node[left,scale=0.8, black] {$\alpha$} (84);
	\draw[alg] (96) -- node[left,scale=0.8, black] {$\alpha$} (85);
	\draw[optalg] (97) -- node[left,scale=0.8, black] {$\alpha$} (86);

	\end{tikzpicture}
	
	\caption{Every edge is announced with $[1,\alpha]$ and revealed with the labeled weight. The adaptive exploration algorithm will traverse on the broad edges, and will not take the small ones (so every second layer). An alternative solution would take the continuous ones (and takes some edges twice in the bottom layer), and ignore the dashed ones. The adaptive exploration algorithm therefore walks on a Hamiltonian path, where each edge costs $\alpha$. The alternative solution only pays $1$ for the majority of the edges in the center, but needs more edges.}\label{graph_lb_grid}
\end{figure}
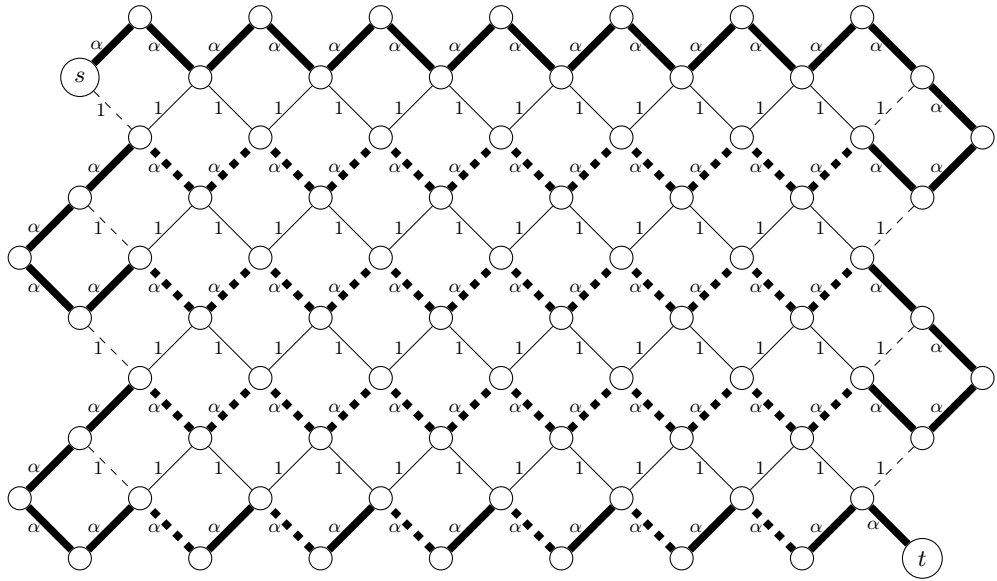

Assuming the algorithm starts at the marked starting vertex $s$ in the grid as depicted in Figure~\ref{graph_lb_grid}. As there exists a Hamiltonian path between the starting vertex and the end vertex $t$, at each point the algorithm prefers staying on the Hamiltionian path instead of visiting a vertex twice just to take a cheaper edge, as taking an edge with weight $\alpha$ is cheaper than taking two edges even with weight $1$. Also traversing between two connected unvisited vertices will cost either the unrevealed edge of at most $\alpha$, or in the best case two revealed edges of weight $1$ as well. Therefore, at every step the algorithm will take the cheapest outgoing edge that is part of a Hamiltonian path.

As in every step of the algorithm (broad path in Figure~\ref{graph_lb_grid}) there is either just one unvisited neighbor that needs to be visited in order to stay on a Hamiltonian path, or one neighbor of which is adjacent to just one unvisited vertex - which also needs to be visited to stay on a Hamiltonian path. Therefore, the adaptive exploration algorithm will traverse the grid in the predetermined path, even if all of those edges are expensive with weight $\alpha$. Note that this also seems to be the strategy which promises the cheapest path through the graph.

On the other hand, an alternative offline algorithm may ignore the fact that a Hamiltonian path seems cheaper for the moment. It also needs to visit the upper and lower row of the grid, thus taking edges with costs of $\alpha$ twice for each vertex in those rows - but it can decide to take the other edges for the middle part of the grid, thus paying only $1$ for each edge but at the cost of taking additional edges for another traversal from the right to the left (depicted by the continuous line).

Overall, for a quadratic grid of $n$ vertices, the adaptive exploration algorithm pays $(n-1)\alpha$, while the depicted alternative solution only pays $6\sqrt{n}\alpha + (\sqrt{n}-2)\sqrt{n}*1$.  Therefore, the competitive ratio between both solutions converges to $\alpha$, as $n$ gets large.
\qed
\end{proof}

The fragility of the adaptive exploration algorithm on instances like a lattice is not surprising, since it allows the adversary to develop traps. Even though it follows a walk that - based on his current beliefs - seems like an algorithm's best hope, it does not value revealing edge weights. If the adversary needs to reveal many edges early, it prevents him from hiding edge weights in order to decide later whether they become cheap or expensive.

\subsection{Outlook on other graph classes}

Many lower bound constructions for the classical graph exploration problem make use of dead ends or the option to either present edges or not depending on the algorithm's actions. This is not possible in the graph exploration problem with edge weight estimates, thus many lower bound constructions of the graph exploration problem will not work here - even for a large factor of $\alpha$.

For example, Miyazaki et al.~\cite{Miyazaki2009TheOG} proved a matching upper and lower bound of $\frac{1+\sqrt{3}}{2}$ for the graph exploration problem on a cycle graph.  For a sufficiently large $\alpha$, the lower bound construction will also work out in our model with edge weight estimates, as there is no structure to hide in this graph class.

But for most graph classes, like tadpole graphs (a cycle with one attached path), the question whether a vertex is on the cycle or on the path is hidden in the classical graph exploration model. Miyazaki et al. \cite{Miyazaki2009TheOG} and Brandt et al. \cite{BRANDT2020176} also showed that even on unweighted tadpole graphs there cannot be an algorithm for the graph exploration problem yielding a better competitive ratio than $2$. This construction cannot work out for the online graph exploration problem with edge weight estimates, as it is clear which edges belong to the path and thus need to be traversed exactly twice by every reasonable algorithm, thus every edge of the path potentially only decreases the competitive ratio.
Therefore, the best possible lower bound for a given $\alpha$ on tadpole graphs is exactly the same construction as for the same factor $\alpha$ on a cycle, and will not exceed a competitive ratio of $\frac{1+\sqrt{3}}{2}$.

\section{Final Remarks and Open Problems}
In this article, we introduced the concept of estimates to the edge weights of the travelling salesman problem resp. graph exploration problem. We believe that the concept is worth looking into: It is not only a natural variant of an exploration problem which can easily be motivated by practical scenarios, but also promises interesting results from the theoretical side.

Since a general solution for the graph exploration problem with edge weight estimates would solve the heavily investigated and long-standing open problem of graph exploration as a special case when the range between the lower- and upper bound gets large, we assume that such a general solution is out of reach with current state-of-the-art approaches. 

Apart from the general question, it is likely fruitful to attack special cases of the graph exploration problem with edge weight estimates. This will hopefully give a better understanding also of the graph exploration problem itself, and might help to come up with a solution in the future. An interesting special case would be the setting with uniform announced edge weights, where we conjecture that there is an $\frac{1+\alpha}{2}$-competitive algorithm for every graph structure and $\alpha \leq 2$.

\begin{credits}
\subsubsection{\ackname} This study was partially supported by the STACS 2024 Extended Stay Support Scheme, funded by the LaBRI, by the NSTC-CNRS International Emerging Action (IEA) 2024-2025 ``Quasi-Polynomial Time Approximation Algorithms for Some Network Design Optimization Problems'' under grants 113--2927--I--141--501 and 114-2927-I-141-501, and by the ANR project TEMPOGRAL (ANR-22-CE48-0001).
\end{credits}

\bibliographystyle{splncs04}
\bibliography{bibliography}

\end{document}